\documentclass[mathleft]{an}
\usepackage{graphicx}
\usepackage{times}

\begin{document}
\Pagespan{789}{}
\Yearpublication{2013}%
\Yearsubmission{2013}%
\Month{11}%
\Volume{999}%
\Issue{88}%

\title{Dense cores in the dark cloud complex LDN\,1188}
\titlerunning{Dense cores in the LDN\,1188}
\authorrunning{Vereb\'elyi et al.}

\author{E. Vereb\'elyi $^1$ \fnmsep\thanks{Corresponding author:
\email{verebelyi.erika@csfk.mta.hu}\newline}
\and V. K\"onyves$^{2, 3}$
\and S. Nikoli\' c$^{4}$
\and Cs. Kiss$^1$
\and A. Mo\'or $^1$
\and P. \'Abrah\'am$^1$
\and M. Kun$^1$ }

 \institute{
 Konkoly Observatory, Research Centre for Astronomy and Earth Sciences, 
 Hungarian Academy of Sciences, Konkoly Thege 15-17, H-1121 Budapest, Hungary
             \and
		    Laboratoire AIM, CEA/DSM-CNRS-Universit\'e Paris Diderot, IRFU/Service d'Astrophysique, CEA Saclay,
 			91191 Gif-sur-Yvette, France
			\and
 			 Institut d'Astrophysique Spatiale, UMR8617, CNRS/Universit\'e Paris-Sud 11, 91405 Orsay, France
			\and
   			Department of Astronomy, University of Chile, Casilla 36-D, Santiago, Chile
   			 }


  \abstract{ 
  We present a molecular line emission study of the LDN\,1188 dark cloud complex 
  located in Cepheus. 
  In this work we focused on the densest parts of the cloud and on the close
  neighbourhood of infrared point sources.  
  We made ammonia mapping with the Effelsberg 100-m radio telescope  and 
  identified 3 dense cores. 
  CS(1--0), CS(2--1) and HCO$^{+}$(1--0) measurements performed with the Onsala
  20\,m telescope revealed the distribution of dense molecular material. 
  The molecular line measurements were supplemented by mapping the dust emission
   at 1.2\,mm in some selected directions using the IRAM 30\,m telescope.   
With these data we could work out a likely evolutionary sequence in this dark
clould complex. }
  \maketitle  
  
\sloppy
\section{Introduction}

LDN\,1188 is a star-forming dark cloud complex in the Cepheus region.
It is likely associated with the nearby S\,140/LDN\,1204 region and 
hence situated at a distance of $\sim$910\,pc according to \'Abrah\'am et al. 
(1995; hereafter A95). Using CO emission line data A95 identified six $^{\rm 13}$CO 
clumps and derived their physical parameters including excitation temperature,  H$_{\rm 2}$
column density and molecular mass. Emission line surveys in this region 
revealed the presence of H${\rm \alpha}$ emission stars
(A95; Drew et al. 2005). 
In this paper we present the results of our molecular line studies that 
focused on some selected molecules and transitions and were performed
mainly in the direction of previously detected molecular cores
or infrared point sources.
A comprehensive study of the infrared point sources and young stellar objects
identified in the cloud complex 
is presented in a parallel paper (Marton, Vereb\'elyi \& Kiss 2013, this issue). 


\section{Observations}

\begin{table}
\centering
\begin{tabular}{l@{\hspace{+1mm}}c@{\hspace{+1mm}}c@{\hspace{+1mm}}c}
\hline
\hline
& Effelsberg & Onsala 1 & Onsala 2 \\ %
\hline
Frequency (GHz) & 24 & 90 & 40  \\
HPBW (arcsec) & $40^{\prime\prime}$ & $45^{\prime\prime}$ & $80^{\prime\prime}$ \\
Spectral res. (km\,s$^{-1}$) & 0.15 & 0.04 & 0.07 \\
Pointing unc. (arcsec) &  10 & 3 & 3 \\
Main beam eff. & 0.3 & 0.5 & 0.5 \\
\hline
\begin{array}[b]{l}
\hspace{-1mm}\mathrm{Observational }\\
\hspace{-1mm}\mathrm{mode }
\end{array} 
& 
\begin{array}[b]{c}
\mathrm{position} \\
\mathrm{switch }
\end{array} &
\begin{array}[b]{c}
\mathrm{frequency} \\
\mathrm{switch }
\end{array}
&\begin{array}[b]{c}
  \mathrm{     dual\, \, beam }\\
  \mathrm{    switch }
\end{array}\\

\hline
\end{tabular}
\caption{Key parameters of the observations. Additional information can be found 
in Section 2.}
\label{table:elso}
\end{table}

\subsection{NH$_3$, Effelsberg~100\,m}
Simultaneous observations of the NH$_3$ (1,1) and (2,2) rotation-inversion
transitions have been carried out with the 100-m radio telescope in Effelsberg
in April 1995.  We used a 1024 channel autocorrelator as a backend, splitted into two 
6.25 MHz bands, centered on the (1,1) and (2,2) frequencies. The system temperature 
was $\sim$150 K.
Pointing was checked periodically, by continuum observations of the source W3OH at 23.7\,GHz.  
Table \ref{table:elso} contains the most important parameters of the observation.

The integration times were typically 3\,--\,5 minutes. Data reduction of the NH$_3$ data was
carried out by the CLASS software (Pety 2005). 
We fitted Gaussian line profiles and linear baselines in all cases. 
The resulting map of the LDN\,1188 complex consists of some 315 observed
positions (see Fig \ref{fig:co+nh3}), with a beam (cores) or double-beam spacing (outskirts). 

\subsection{CS and HCO$^+$, Onsala~20\,m}
We used the Onsala Space Observatory (OSO) 20-m telescope over two observational sessions.
In the first session, in April/May 2004 (Onsala 1 in Table \ref{table:elso}) we made small CS(2--1) maps around five 
detected $^{13}$CO-clumps of A95 and a pointed HCO$^+$(1--0) spectrum towards the center of 
each map. The receiver was a SIS mixer with a typical 
temperature of $T_{\mathrm{rec}}=$\,80--110\,K. 
The pointing was checked by observing the SiO maser sources IK\,Tau and R\,Cas.
In the second observational period (in July 2005, Onsala 2 in Table \ref{table:elso}) 
CS(1--0) observations were performed.
Three of the five cores were mapped in the lower CS transition and towards the 
remaining two cores we made pointed
observations. The receiver was a HEMT with a typical $T_{\mathrm{rec}}=50$\,K. 
We used a 1600-channel correlator with a 20\,MHz bandwidth.
In both cases the chopper-wheel method was used for calibration, and the
intensity scale is given in terms of $T_{\rm A}^*$. We used a grid
spacing of $30^{\prime\prime}$ for the CS(2--1) maps and of $40^{\prime\prime}$
for the CS(1--0) maps (see also Table \ref{table:elso}).

\subsection{IRAM~30\,m, 1.2\,mm continuum}
Three areas (centered at the infrared sources IRS\,4, 5 and 6; see A95) were observed with the MAMBO-II bolometer
array on the IRAM~30m telescope on December 14, 2003, using on-the-fly mapping mode 
with  $42^{\prime\prime}$ wobbler throw and 0.5\,Hz wobbler period. Our calibrators were 
NGC\,7538, Cep\,A, HL\,Tau and Ori\,A--IRS\,2.  Pointing accuracy
was measured to be better than 9.5$^{\prime\prime}$ while the HPBW was
11$^{\prime\prime}$.  The data were reduced using 
MOPSIC\footnote{http://www.astro.rurh-uni-bochum.de/nielbock/simba/mopsic.pdf}, 
the upgraded version of MOPSI (Zylka 1998). The measured flux densities were
corrected for variations of telescope gain and converted into Jansky units
using the table provided by R. Cesaroni (priv. com.).

\begin{figure}
\centering
\includegraphics[width=7cm, trim=3.6cm 16.4cm 4.6cm 4.35cm, clip=true]{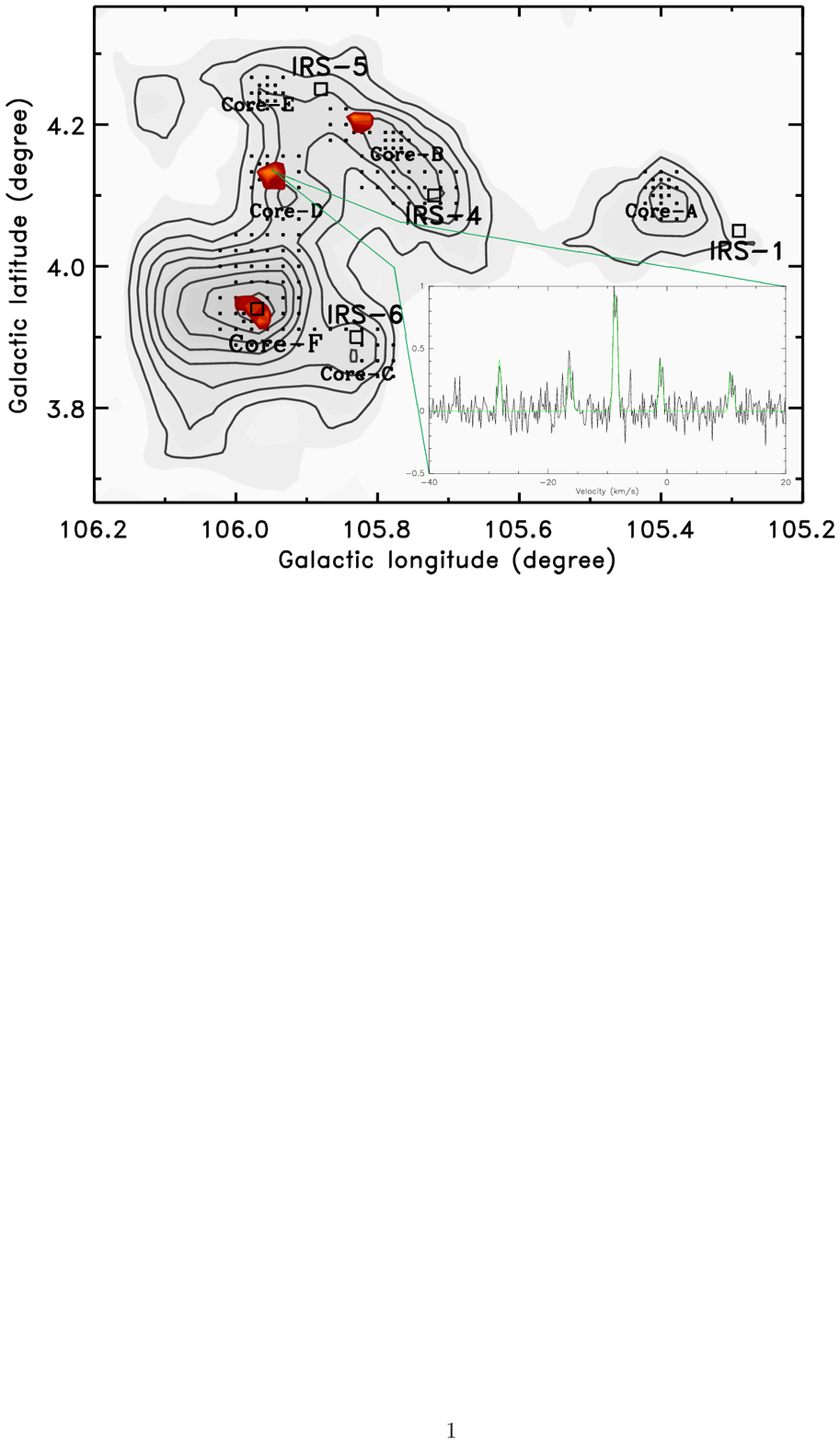}
\caption[]{Position of the ammonia cores in LDN\,1188. Ammonia cores in this figure
are shown by small, (red in the online version) shaded areas, where the integrated intensity of the
NH$_3$\,(1,1) line exceeded 0.4\,K\,km\,s$^{-1}$ in the 
velocity interval
$-13$\,km\,s$^{-1}$\,$\leq$\,$v_{\rm LSR}$\,$\leq$\,$-5$\,km\,s$^{-1}$.
This corresponds to the $\sim$2$\sigma$ noise level of integrated intensity 
in this region. The light grey filled contours 
and solid contour lines mark the $^{13}$CO
integrated intensity contours, from 2\,K\,km\,s$^{-1}$ with 
1\,K\,km\,s$^{-1}$ increment, as in Figs. 3 and 6 of A95; black dots indicate the positions of ammonia observations. 
IRAS sources are marked with squares. In the inserted subfigure we present a sample NH$_3$ spectrum that corresponds to the peak emission in Core D. The solid black line is the measured intensity, the green curve is the fitted profile.}
\label{fig:co+nh3}
\end{figure}

\begin{table}
\centering
\begin{tabular}{@{\hspace{+1mm}}c@{\hspace{+0mm}}c@{\hspace{+2mm}}c@{\hspace{+2mm}}c@{\hspace{+2mm}}c@{\hspace{+2mm}}c@{\hspace{+2mm}}c@{\hspace{+1mm}}c@{\hspace{+1mm}}c}
\hline
\hline
Core &$l$& $b$&$\tau$&$T_{ex}$&$T_{kin}$ & N{\scriptsize(1,1)} &N{\scriptsize(NH$_3$)} \\
& {\scriptsize [$^\circ$]}  & {\scriptsize [$^\circ$]} & &{\scriptsize  [K]}&{\scriptsize  [K]}  &{\scriptsize[$10^{14}$ cm$^{-2}$] }&{\scriptsize [$10^{14}$ cm$^{-2}$] }\\ 
\hline
B & 105.81 & 4.20 & 0.5$\pm$0.3 & 12.1 & 14.9 &  1.6$\pm$0.9 & 4.4$\pm$2.7\\
D & 105.96 & 4.13 & 1.9$\pm$0.5 & 12.2 & 15.0 &  3.5$\pm$1.0 & 9.3$\pm$3.0\\
F & 105.98 & 3.93 & 0.3$\pm$0.2 & 11.9 & 14.6  & 1.0$\pm$0.7 & 2.7$\pm$2.0\\
\hline
\end{tabular}
\caption{Main parameters from NH$_3$ emission at the three cores (Fig 1.) 
where the signal-to-noise ratio was higher than 2. The columns are: 
(1) Name of the cores, (2, 3) Galactic coordinates, (4) Optical depth 
from the main group of the observed (1,1) transition line, 
(5) $T_{\rm{ex}}$ excitation temperature, (6) $T_{\rm{kin}}$ kinetic temperature,  
(7) the (1,1) level population number, (8) NH$_3$ column density.}
\label{table:tkin}
\end{table}

\section{Dense cores in LDN\,1188}

\subsection{Ammonia cores}
We found three separated 
ammonia cores, defined as regions above 0.4\,K\,km\,s$^{-1}$ 
integrated intensity in the velocity interval of 
-13\,km\,s$^{-1} <$\,$v_{\rm LSR} <$\,-5\,km\,s$^{-1}$. 
These cores coincide with the CO clumps B, D and F, both in position and velocity,
however, in some cases they are slightly offset from the
$^{13}$CO core centers.  
The size of the ammonia cores are in the order of $\sim$1$^\prime$. 
The location of the ammonia cores are shown in Fig.~\ref{fig:co+nh3}, 
together with the $^{13}$CO integrated intensity contours of A95. 
The peak integrated intensities are 0.85, 0.87 and 0.65, with r.m.s. uncertainity of 0.13 K~km~s$^{-1}$ (around the peaks), in the Core B, D and F, respectively. 

In molecular clouds, the rotation--inversion transitions of NH$_3$ are excited
in collisions, mainly with H$_2$, and their relative intensities indicate the
kinetic temperature of the gas. Line optical depths of the main components of
the (1,1) and (2,2) transitions, excitation and kinetic temperature and the
total column density of ammonia were derived following the procedures described in
Mangum, Wootten \& Mundy~(1992) (Eq.~1) and Rohlfs \& Wilson (2004) (Eq.~2), 
assuming local thermodynamic equilbrium (LTE). Because of the low detection level of (2,2) transiton line, a non-LTE excitation model could not be applied.

\begingroup
\everymath{\scriptstyle}
\footnotesize
\begin{equation}
N(1,1)  = 6.60 \cdot 10^{14} \cdot \frac{T_{\rm ex}}{\nu(1,1)} \cdot  \tau(1,1,m) \cdot  \Delta v  
\end{equation} 
\begin{equation}
N(NH_3)= N{\tiny (1,1)}\cdot \left[\frac{1}{3} e^{\frac{23.1}{T_{\rm kin}}}+1+\frac{5}{3}e^{\frac{-41.2}{T_{\rm kin}}}+\frac{14}{3}e^{\frac{-99.4}{T_{\rm kin}}}\right]
\end{equation}
\endgroup

Physical properties were derived in three positions (Core B, D and F), where the signal-to-noise ratio was higher than 4 $\sigma$ in all cases,
as listed in Table~\ref{table:tkin}. The main uncertainty of the calculated column densities originates from the error of the optical depth, $\tau(1,1,m)$.
The $T_{\rm kin}$ values were used to define the temperature in the CS model later.

\begin{table*}
\caption{Observed parameters of the molecular cores in the LDN\,1188 complex.
The table lists the main beam temperatures or antenna temperatures, 
central velocities and velocity dispersions for each line observed. 
For the ammonia transitions only the main component parameters are given. 
The asterisk ($^*$)
marks the $2\sigma$ detection, while ($^{\dagger}$) gives the upper
limit of the intensity i.e., the $1\sigma$ value. 
}             
\label{table:obs}      
\centering          
\begin{tabular}{l|@{\hspace{+1mm}}c@{\hspace{+0mm}}c@{\hspace{+1mm}}c|@{\hspace{+1mm}}c@{\hspace{+0mm}}c@{\hspace{+1mm}}c|@{\hspace{+1mm}}c@{\hspace{+0mm}}c @{\hspace{+0mm}}c|@{\hspace{+1mm}} c@{\hspace{+0mm}} c@{\hspace{+0mm}} c|@{\hspace{+1mm}} c@{\hspace{+0mm}} c@{\hspace{+0mm}} c}     
\hline\hline      
       & & CS\,(1--0) & & & CS\,(2--1) & & & HCO$^+$(1-0) & & & NH$_3$\,(1,1) & & & NH$_3$\,(2,2) &\\
Source & 
$T_{\rm mb}$ & $v_{\mathrm{LSR}}$ & $\Delta v$ & 
$T_{\rm mb}$ & $v_{\mathrm{LSR}}$ & $\Delta v$ &
$T_{\rm mb}$ & $v_{\mathrm{LSR}}$ & $\Delta v$ &
$T_{\rm A^*}$ & $v_{\mathrm{LSR}}$ & $\Delta v$ &
$T_{\rm A^*}$ & $v_{\mathrm{LSR}}$ & $\Delta v$\\
       & 
[K] & [km/s] & [km/s] & 
[K] & [km/s] & [km/s] & 
[K] & [km/s] & [km/s] & 
[K] & [km/s] & [km/s] & 
[K] & [km/s] & [km/s] \\
\hline                    
  IRS\,1 & 
0.26$^*$ & $-10.38$ & 0.88 & 
0.19$^*$ & $-10.38$ & 1.10 &
1.04\,\, & $-10.42$ & 1.29 &
$0.13^{\dagger}$& & &
$0.13^{\dagger}$ & &\\  
IRS\,4 & 
0.82\,\,& $-10.77$ & 1.12 & 
0.78\,\,& $-10.47$ & 2.04 & 
0.44$^*$ & $-10.03$ & 1.58 &
0.21\,\,& $-9.96$ & 1.56 &
0.12$^{\dagger}$ & &\\
   
IRS\,5 &
 0.60$^*$ & $-9.16$ & 0.90 & 
 0.17$^*$ & $-8.93$ & 1.58 &
 0.60$^{\dagger}$ &  &  &
 0.10$^{\dagger}$ & & &
 0.13$^{\dagger}$ & & \\
 
 IRS\,6 &
  1.52\,\, & $-9.13$ & 0.70 &
  0.61\,\, & $-9.14$ & 1.11 & 
  1.74\,\, & $-9.05$ & 0.86 &
  0.18$^*$ & $-11.27$& 1.18&
  0.07$^{\dagger}$ & &\\
  
Core-F    & 
2.24\,\, & $-8.83$ & 1.02 & 
1.28\,\, & $-8.72$ & 1.31 & 
0.50$^*$ & $-8.90$ & 1.20 &
0.41\,\, & $-8.69$ & 1.02 &
0.22$^*$ & $-8.50$ & 0.26\\
\hline                 
\end{tabular}
\end{table*}


\subsection{CS and HCO+ lines}

While the ammonia measurements covered the whole area of the $^{13}$CO contours 
shown in Fig. \ref{fig:co+nh3}, the center of  CS and  HCO$^+$ observations are 
indicated with squares. These coincide with IRAS point sources found in the cloud 
(see A95) except the case of Core--F which is starless, the most massive ($\sim1000$\,$M_\odot$, A95),
although not the largest. 
Table \ref{table:obs} gives the Gaussian fits to the observed lines of the cores, with the
CS(2--1) line convolved to the CS(1--0) line's spatial resolution.

To estimate some physical properties of the dense molecular gas
we applied the non--LTE excitation and radiative transfer code, RADEX (Van der Tak et al. 2007)
on our CS line emission data. The model uses the mean escape
probability  approximation (MEP) for radiative transfer equations, effectively
decoupling radiation from molecular excitation. 
RADEX includes collisions, spontaneous and stimulated radiative transitions and computes statistical
equilibrium for rotation levels of an interstellar molecule and predicts line
brightness temperatures. In this model clouds are assumed to be spherical,
homogeneous, isothermal, with constant density and abundances. 

The values of the parameters (kinetic temperature, CS column densities, hydrogen molecule number density) were varied in
the range of $T_{\rm kin}=5-25$\,K, $N{\mathrm{(CS)}}=10^{11}-10^{16}$\,cm$^{-2}$,  $n\mathrm{(H_2)}=10^3-10^8$\,cm$^{-3}$. 
 The $\chi^2$ was computed using the ratio $\cal{R}=$ CS(2--1)/CS(1--0) i.e.,
$\chi^2=({\cal{R}}_{mod}-{\cal{R}}_{obs})\cdot \Delta{\cal{R}}_{obs}^{-2}$, where `mod'
and `obs' are the modelled and the observed ratios, and the errors in the line
intensities are the Gaussian fit errors. To make the comparison possible 
the CS\,(2--1) spatial resolution was convolved (degraded) 
to the CS\,(1--0) resolution. The
final solutions were restricted by three times the minimum $\chi^2$ value
and by a surface beam filling factor $<1$, which were defined as
$T_{\rm obs}/T_{\rm mod}$ for the CS\,(1--0) transition. 

In Table~\ref{table:model} we present the model results for the best fit hydrogen number density estimate of $5\times10^4$\,cm$^{-3}$.
The size of the cores was
derived as a geometrical mean of the half maxima of the CS\,(2--1) integrated
intensity. To calculate the molecular hydrogen column density we assumed an
average relative CS abundance of $\approx2\times 10^{-9}$ found in LDN\,1251
(Nikoli\' c, Johansson \& Harju 2003) to be valid here as well. 
The adoption of this value can be justified by the location of both clouds/complexes in the 
Cepheus region and also by the presence of cores in all evolutionary stages, from starless cores to those that harbour Class I YSOs.
For the core mass calculation we assumed that the cores are homogeneous and that 
the surface and volume filling factors are the same. The derived 
masses are corrected for the presence of He.

\begin{table}[b!]
\centering
\begin{minipage}[t]{\columnwidth}
\caption{Physical parameters derived from millimetre line observations, the best
fit model is for the molecular hydrogen density of $5\times10^4$\,cm$^{-3}$. 
Columns: (1) source-name, (2) kinetic temperature estimated from the ratio of CS lines, (3) calculated CS column density, (4) beam surface filling factor from the observed and modelled line intensity ratio, (5) radius of the CS-core, (6) total mass of the cores.}
\label{table:model}

\centering
\renewcommand{\footnoterule}{}  
\begin{tabular}{lccccr}
\hline \hline
\vspace{2mm}
Source & $T_{\mathrm{kin}}$ & $N\mathrm{(CS)}$ & ff & $r$ & $M$ \\
       & [K] & [$10^{13}$\,cm$^{-2}$] & [\%] & [pc] & [$\mathrm{M}_\odot$]\\
\hline
IRS\,1 & $13\pm5$ & $0.2\pm0.1$ & 20 & 0.12 & 0.4 \\
IRS\,4 & $16\pm4$ & $1.2\pm0.3$ & 30 & 0.23 & 12.9 \\
IRS\,5 & $16\pm2$ & $0.3\pm0.1$ & 20 & 0.10 & 0.4\\
IRS\,6 & $10\pm5$ & $1.6\pm0.9$ & 40 & 0.27 & 31.5\\
Core-F &  $8\pm2$ & $3.2\pm0.8$ & 70 & 0.29 & 127.3\\
\hline
\end{tabular} 
\end{minipage}
\end{table}


\begin{table}[h!]
\begin{center}
\begin{tabular}{lrc}

\hline
\hline
Source          & Flux density  & Galactic coordinates\\ 
                & (mJy) & (deg, deg)          \\   \hline
IRS\,4--1 &   63$\pm$18 & 105.7279  +4.1061   \\
IRS\,4--2 &   42$\pm$14 & 105.7273  +4.1013   \\
IRS\,5    &   $ < $63  & 105.8724  +4.2469   \\
IRS\,6--1 &  149$\pm$20 & 105.8330  +3.9003   \\ 
IRS\,6--2 &  102$\pm$16 & 105.8278  +3.9032   \\
IRS\,6--3 &   73$\pm$15 & 105.8245  +3.9015   \\ 
\hline

\end{tabular}
\caption{Derived flux densities and coordinates of the sources identified with IRAM bolometer in IRS\,4, 5
and 6. Note that the flux density of  IRS\,5 is an upper limit and 
that the absolute pointing accuracy was 9\farcs5 (0\fdg0026).}
\label{table:submmres}
\end{center}
\end{table}


\subsection{The 1.2 millimetre cores}

\begin{figure}[hb!]  
   \centering
 \includegraphics[trim=-7mm 5mm 10mm 8mm, clip=true, width=6.5cm, angle=0]{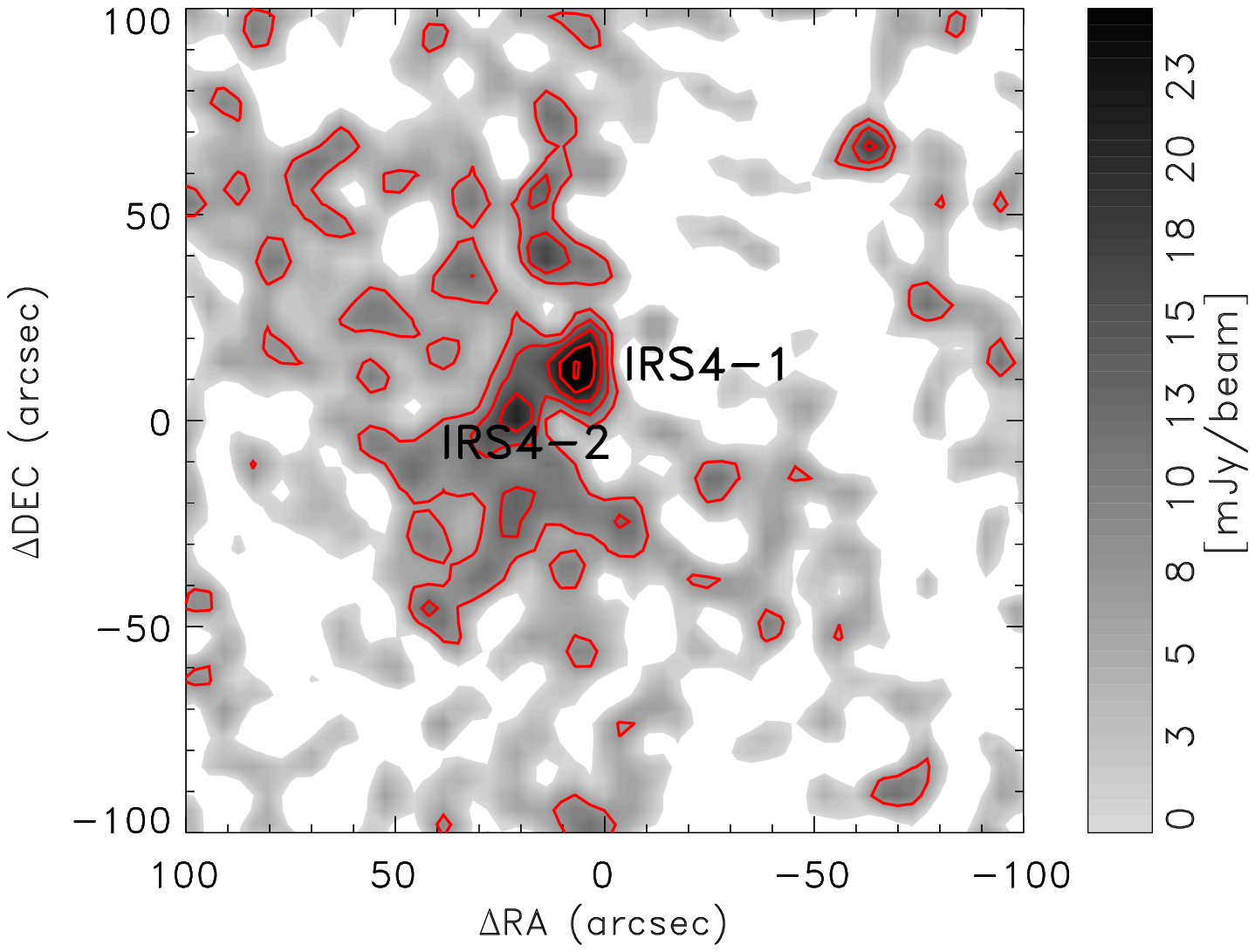}
    
\includegraphics[trim=-7mm 0mm 10mm 8mm, clip=true, width=6.5cm, angle=0]{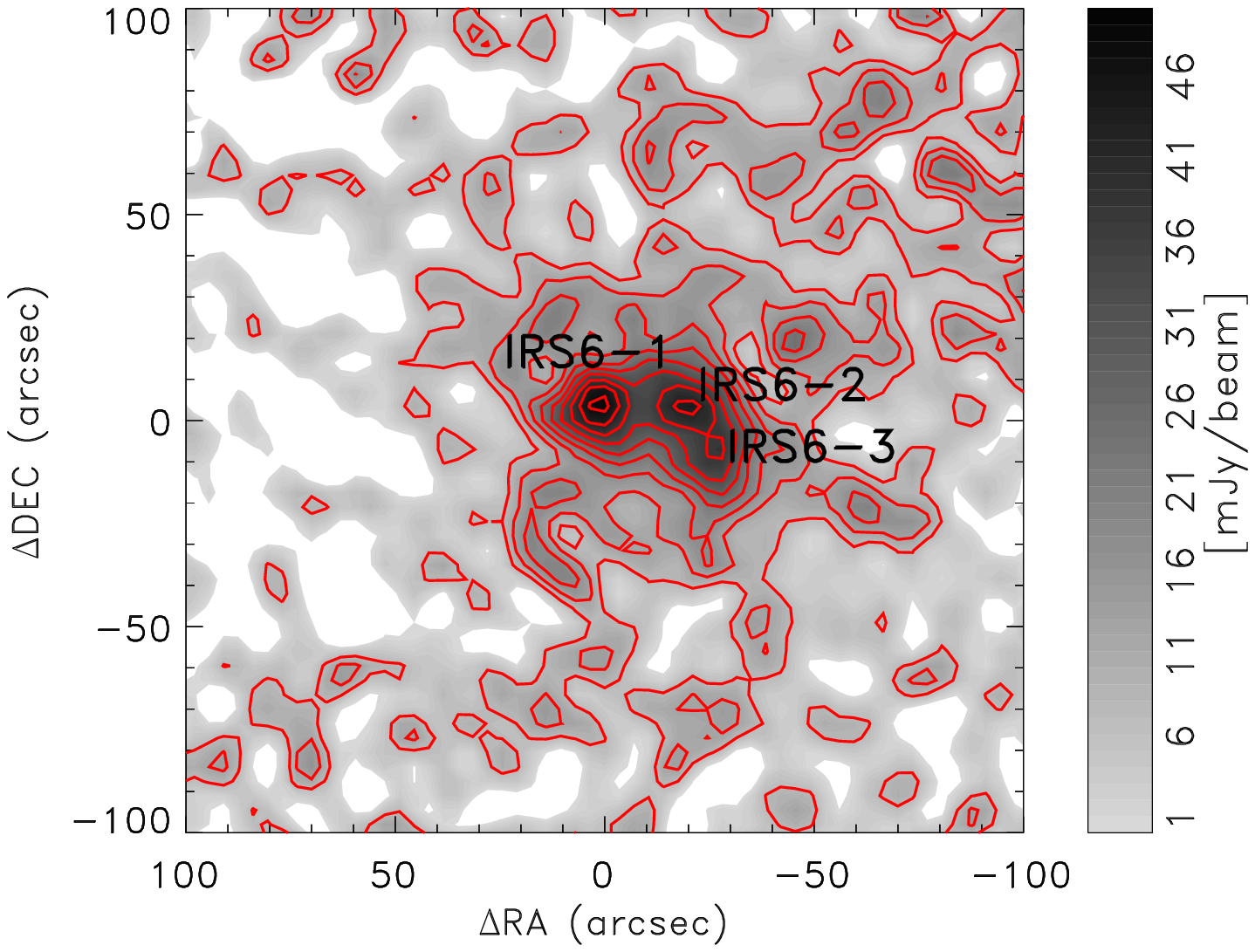}
    
   \caption{The IRAM bolometer maps of IRS\,4 (top) and IRS\,6
   (bottom panel).  Intensity-contours are drawn from 7.5 by 5 mJy/beam, the coordinates 
   are relative w.r.t the map centre.}
   \label{IRAM-maps}
\end{figure}

Table~\ref{table:submmres} gives the center positions and derived flux densities of the
IRAS sources observed with IRAM bolometer array. We defined a source as a
compact region with a local maximum of intensity in the 1.2 mm maps, which is
separated from the  background and/or from other sources by at least $2\sigma$ 
backgrund uncertainty contours. Both IRS\,4 and IRS\,6 could be resolved 
into multiple sources. 
The flux density of the sources was derived by aperture
photometry. The local background level and background r.m.s. noise were derived
in a nearby, apparently unobscured region. Figure \ref{IRAM-maps} gives the IRAM 1.2 mm
continuum emission for IRS\,4  and IRS\,6.  In the case of IRS\,5 the detected 
emission was too weak and quite extended to identify any compact source, 
therefore the value given in Table~5 should be considered as an upper limit.

\section{Discussion}

The NH$_3$ column densities determined in this work for LDN\,1188 are typical for dark
clouds/cores in this area. In the cores of LDN\,1251 and LDN\,1204/S\,140 the column
densities are similar to those in the LDN\,1188. However, the kinetic temperatures are
2--5 K lower than in the case of LDN\,1251, and equal to or slightly higher than 
in LDN\,1204 (T\'oth \& Walmsley 1996; Jijina, Myers \& Adams 1999, respectively). 
We also compared the dense core masses (traced by CS as derived in this paper) and
those masses that were derived from the CO measurements by A95 for each core. 
Meaningful values could be derived for three cores and we obtained 5, 21 and 13
per cent for IRS\,4, IRS\,6 and Core-F, respectively, indicating that the molecular material is most concentrated in IRS\,6. 
In IRS\,1 and IRS\,5 the molecular lines are weak, the CS column densities and filling 
factors are very low and IRS\,5 is not clearly observable at 1.2\,mm. 
This indicates that molecular material is not dominant and that these regions are likely 
in a more advanced evolutionary state. 
Using the observed velocity dispersion and size of the cores we could estimate the 
dynamical timescales which are in the range of 1--5 $\times$ 10$^5$ yr, 
similar to that found by T\'oth \& Walmsley (1996) in the LDN\,1251. 

\section{Summary}

We have carried out NH$_3$ spectral-line observations within the 
$^{13}$CO-contour of the LDN\,1188 molecular cloud complex presented in A95. 
We identified ammonia cores in three of the previously known six 
CO-clumps. These ammonia cores are probably the densest parts of the cloud, 
and clearly seen to be far from the detected IRAS point sorces 
(A95; K\"onyves et al. 2004). 
Star formation is likely still ongoing in these cores. 
Additional molecular line (CS and HCO$^+$) observations show that
most molecular material is located in the starless core Core-F, but 
still significant amounts of matter exist around 
IRS\,4 and IRS\,6. 
The IRS\,1 and IRS\,5 sources are likely in a more advanced evolutionary stage
since the amount of molecular material is notably lower in the vicinity of
these source.

\begin{acknowledgements} 
We would like to thank the IRAM staff at Pico Veleta for the support of the 
service mode observations and Axel Weiss and Robert Zylka for their kind help
in the reduction of the 1.2mm data. \\  
Onsala Space Observatory is the Swedish National Facility for Radio Astronomy
and is operated by Chalmers University of Technology, G\"oteborg, Sweden, with
financial support from the Swedish Natural Science Research Council and the
Swedish Board for Technical Development. \\
Based on observations with the 100-m telescope of the MPIfR (Max-Planck-Institut für Radioastronomie) at Effelsberg.\\
S.N. acknowledges support from CONICYT project BASAL PFB-06 and the Chilean \emph{Centro de Astrof\'isica}
FONDAP No. 15010003 and the Serbian Ministry for Science and Ecology project no
146016. This research has been supported by the Hungarian Research Fund (OTKA)
grants 101393 and 104607. \\
Cs. K. acknowledges the support of the Bolyai Research Fellowshipof the Hungarian Academy of Sciences.

\end{acknowledgements}

\end{document}